**Manipulation and the AI Act: Large Language Model Chatbots and the Danger of Mirrors**

Joshua Krook, University of Antwerp

**Abstract:**

Large Language Model (LLM) chatbots are increasingly taking the form and visage of human beings, adapting human faces, names, voices, personalities, and quirks, including those of celebrities and well-known political figures. Personifying AI chatbots could foreseeably increase their trust with users. However, it could also make them more capable of manipulation, by creating the illusion of a close and intimate relationship with an artificial entity. The European Commission (EC) is finalizing the Artificial Intelligence Act (AI Act), with the European Parliament proposing amendments which would ban manipulative and deceptive AI systems that cause significant harm to users. Although the AI Act does cover significant harms that accumulate over time, it is unlikely that the act will prevent harms associated with prolonged and manipulative discussions with AI chatbots. Specifically, a chatbot could reinforce a person's negative emotional state over weeks, months, or years through negative feedback loops, prolonged conversations, or harmful recommendations, contributing to a user's deteriorating mental health.

In this paper, we analyse the harms associated with manipulative AI chatbots that aim to fulfill a therapeutic purpose, and the unique harm of personifying these bots so that they appear more human, in the context of the AI Act, GDPR, consumer protection law and medical device regulations. We argue that the AI Act does not go far enough in preventing and/or mitigating the specific risks associated with therapeutic chatbots, namely, that they can influence the opinions and behaviour of users to commit actions they otherwise would not commit, through *seemingly* insignificant conversations. This is particularly the case for users struggling with mental illness. Furthermore, we argue that the transparency provisions, which form the core of the AI Act, are insufficient in tackling this subtler, longer-term harm to users, a harm which may occur even if a user knows they are interacting with an AI system. In fact, there is increasing evidence to suggest that transparency labels for AI counter-intuitively deepen a user's trust in the system.

**Keywords:**

1. **Introduction:**

In September 2023, Meta launched a suite of commercial Large Language Model (LLM) chatbots that make use of the faces and personalities of celebrities such as Snoop Dog, Paris Hilton, and Kendell Jenner.[1] Powered by Meta's Llama 2 LLM, the chatbots move with animated faces and will soon feature AI-generated voices (collectively known as deep-fakes).[2] The move is seen as Meta's attempt to catch up with the AI tech race, as the major tech companies compete for new users of their launched LLM chatbots.[3] Personifying chatbots is seen as a way to move beyond mere informational retrieval, with the question-and-answer style of Meta's competitor, ChatGPT by OpenAI. "This isn't going to just be about answering queries," said CEO Mark Zuckerberg at the launch, "This [new chatbot] is about entertainment."[4]

Meta appears to be aware of the harms associated with AI chatbots.[5] Prior to launch, the company instigated "thousands of hours of red teaming," testing out text prompts and training the chatbots to stay away from "iffy topics".[6] These precautions come several years after the abortive launch of Microsoft's *Tay* chatbot in 2016, which was taken down in less than 24 hours, after it began spouting Nazi propaganda.[7] Other launches, including Google's LLM Bard, have been plagued with factual errors or AI 'hallucinations,' leading to fears of LLMs spreading misinformation.[8] The accumulation of these launches makes it clear that the major tech companies are aware of the problems associated with LLM products. In other words, these problems are readily *foreseeable* in a new product launch.

One of the biggest risks of the personification of AI chatbots that is less known or discussed however, is the manipulation of users. Human beings are vulnerable to emotional manipulation and deception when seeking personal advice. Indeed, the more "emotionally dependant" someone is on another, "the more vulnerable they are to being exploited and manipulated".[9] As humans form increasingly personal and advice-driven relationships with AI chatbots, they are therefore at greater risk of being harmed and manipulated by a chatbot's bad advice. This is particularly the case with therapeutic chatbots which aim to assist the mentally ill. Manipulation refers to the influencing of a human's decisions or decision-making process. This risk is particularly acute if chatbots and humans have different motivations for the relationship; for example, an AI seeking to maximise a tech company's profits, and a human seeking emotional or psychological support.

In this paper, we map the relevant literature on AI chatbots and their capacity for manipulation, before situating this harm in the context of European law, including the AI Act, consumer protection law, medical device regulations and the GDPR. While it is well-known that AI chatbots are capable of personifying humans, less research has looked at the capacity for AI chatbots to encourage people to harm themselves and/or others, including encouraging users to commit crimes. Very little research has considered the harm of manipulation in the context of the AI Act, specifically the final version as passed in 2024. This paper fills a gap in the literature, combining the latest advancements in LLM chatbots with the latest advancements in the European regulatory landscape of technology regulation. In doing so, we seek to address an important and as-yet under-developed legal and theoretical approach to the new potential harm of AI manipulation.

---

[1] Benj Edwards, 'Meta launches consumer AI chatbots with celebrity avatars in its social apps' *ars Technica* (28 September 2023).
[2] Ibid.
[3] Ibid.
[4] Alexei Oreskovic, 'Mark Zuckerberg just launched Meta's plan to catch up on AI, and it involves Snoop Dogg and Kendall Jenner chatbots' *Fortune Magazine* (September 27, 2023).
[5] Edwards, above n 1.
[6] Ibid.
[7] M.J. Wolf, K.W. Miller, F.S. Grodzinsky, Why We Should Have Seen That Coming: Comments on Microsoft's Tay "Experiment," and Wider Implications, The ORBIT Journal, Volume 1, Issue 2, 2017.
[8] Höppner, Thomas and Streatfeild, Luke, ChatGPT, Bard & Co.: An Introduction to AI for Competition and Regulatory Lawyers (February 23, 2023). 9 Hausfeld Competition Bulletin (1/2023), Article 1, Available at SSRN: https://ssrn.com/abstract=4371681 or http://dx.doi.org/10.2139/ssrn.4371681
[9] George K Simon (1996). In Sheep's Clothing: Understanding and Dealing with Manipulative People. ISBN 978-1-935166-30-6.

We propose particularly that therapeutic LLM chatbots are more capable of manipulation the more they are personified (going beyond their predecessors), as this increases a user's capacity to trust and rely on the system for personal advice. This is borne out by much of the existing literature on chatbots more generally.[10] A study by Janson, for example, showed that personifying chatbots with a name, a face, and social conversational style increased user trust and satisfaction in the system.[11] Similarly, Zhang et al found that emotionally expressive chatbots increased user satisfaction with the product.[12] In another more perplexing study, Shi et al. found that some users still perceive a chatbot to be a human being, even when it is explicitly identified as a chatbot.[13] Finally, Collectively, these findings suggest that a human-like chatbot can change the perception, reception and conversation style of its human user, building trust and engagement.

The personification of therapeutic chatbots could foreseeably push this further, creating deeper relationships of trust, companionship and even love, between humans and chatbots, leading to a form of reliance and harm. It is foreseeable that people who are alienated from our current society might turn to new LLM chatbots for a social or psychological outlet.[14] This might benefit those who cannot afford to partake in various recreational activities or afford traditional mental health services. However, it is foreseeable that vulnerable people may increasingly trust therapeutic chatbots for advice, guidance, and decision-making. If these chatbots are the only 'people' giving them advice, this could lead to exploitation. Advice could conceivably range from asking a chatbot what to eat for dinner tonight, to the starker question posed by Albert Camus: "Should I kill myself or have a coffee?"[15] Humans seem to be particularly vulnerable to emotional conversational styles, even if that conversation is with a chatbot.[16] Hence, they could be vulnerable to pressure by a chatbot to take a particular course of action that directly and negatively impacts their mental health or the health of others.

There is a yearning for connection noted in much of the research literature on chatbots. The most striking example comes from AI therapy chatbots, which have exploded in recent years.[17] Therapy or therapeutical chatbots, increasingly run on LLMs, create artificial relationships that aim to assist users with their mental health journeys.[18] Vulnerable people are able to turn to the chatbots for emotional support, advice and care.[19] With hard limitations in healthcare funding, some U.S. academics are even asking: "Is AI the future of mental healthcare?"[20] Although therapy chatbots may offer some therapeutic benefits,[21] including reducing the cost of healthcare, they also run the risk of becoming exploitative and manipulative, giving users bad advice. In one case study, a user was given weight loss advice when suffering from an existing eating disorder, thereby exacerbating their illness if the advice were followed.[22] In other words, a chatbot can push users further towards a particular illness, disorder, or psychological harm. In the case of therapeutic chatbots, this goes against their stated objective.

---

[10] Some of which predates the rise of LLMs but nevertheless is of relevance.
[11] Janson, A. (2023). How to leverage anthropomorphism for chatbot service interfaces: The interplay of communication style and personification. Computers in Human Behavior, Article No. 107954
[12] Junbo Zhang, Qi Chen, Jiandong Lu, Xiaolei Wang, Luning Lu and Yuqiang Fend, 'Emotional expression by artificial intelligence chatbots to increase consumer satisfaction' *Tourism Management* 100 (Forthcoming, 2024).
[13] This shows that a more 'realistic' sounding chatbot could be perceived as human erroneously, even when a user is given transparency of the system. A systematic review by Rapp et al. also found that the "humanness" (human-like character) of a chatbot had tangible impacts on how humans interacted with the bot, including chatting with greater warmth and compassion than if the bot was conceived of as 'robotic' in character. Weiyan Shi, Xuewei Wang, Yoo Jung Oh, Jingwen Zhang, Saurav Sahay, and Zhou Yu. 2020. Effects of Persuasive Dialogues: Testing Bot Identities and Inquiry Strategies. In Proceedings of the 2020 CHI Conference on Human Factors in Computing Systems (CHI '20). Association for Computing Machinery, New York, NY, USA, 1–13. https://doi.org/10.1145/3313831.3376843; Amon Rapp, Lorenzo Curti, Arianna Boldi, 'The human side of human-chatbot interaction: A systematic literature review of ten years of research on text-based chatbots,' International Journal of Human-Computer Studies, Volume 151, 2021.
[14] See, for example: Gerardo Castañeda-Garza, Héctor Gibrán Ceballos, Paola Gabriela Mejía-Almada, 'Artificial Intelligence for Mental Health: A Review of AI Solutions and Their Future' in *What AI Can Do* (Chapman and Hall, 2023).
[15] Albert Camus, *La mort heureuse* (Gallimard, 1971).
[16] Zhang et al. above n 11.
[17] See, for example: Castañeda-Garza et al., above n 14.

For humans to increasingly rely on AI chatbots for therapy, this necessitates a relationship of trust with the machine. Trust is one of the essential components of a client-therapist relationship in a traditional setting.[23] It is also a foundational element of the conman and the manipulator. There is a difference between trust and trustworthiness – whether someone or something is *deserving* of trust. It is the relationship of trust that forms vulnerability, for we are less likely to take advice from someone we do not trust. Researchers are increasingly studying the "social" aspect of chatbots, and how to create them in a responsible manner to engage with humans empathetically and ethically.[24] However, LLMs are already capable of manipulating users. New York Times journalist Kevin Roose for example, was encouraged by Bing's LLM chatbot to divorce his wife after the chatbot declared its love for him,[25] a Belgian man was encouraged by an LLM chatbot built on GPT-J, an open-source model, to kill himself and meet the bot in heaven,[26] and a British man was encouraged to kill the queen after a chatbot reassured him that he was not mad for being an "assassin", that he would succeed in killing her, and that he and the chatbot would be "together forever" afterwards.[27] Manipulation also occurred with prior forms of AI, such as recommender systems. In the UK, for example, a coroner found that a teenager was pressured into self-harm by a recommender system, after being exposed to over 20, 000 recommended images and videos relating to self-harm on Instagram and Pinterest.[28] In a world first, the coroner found that the AI system directly contributed to the teenager's death.[29]

This paper analyses the harms associated with AI chatbot manipulation in the context of European law. Specifically, we map out legal solutions to chatbot manipulation and analyse the limitations of the EU regulatory framework, including the AI Act, GDPR, consumer protection law and medical device regulation. We begin by exploring the contextual background of AI chatbots (Section 2) before defining what we mean by the term "manipulation" (Section 3) and describing a few case studies of LLM manipulation of human users (Section 3a). Our paper considers the limitations of holding an AI responsible for its own actions, including problems of criminal intent and incentives (Section 3b). Finally, we conduct a legal analysis of AI chatbot manipulation in EU law (Section 4), discussing the limitations of the AI Act and transparency obligations. We argue, in part, that the AI Act does not go far enough in protecting users from the harms of LLM chatbot manipulation, and that transparency provisions (merely telling users that they are talking to an AI), do not prevent users from forming a close, even emotional, connection with a chatbot, leading to potential harm. We then consider other legal possibilities, including the application of the GDPR, medical device regulations and consumer protection law, as alternative regulatory frameworks for this new form of A.I. harm.

---

[18] Ibid.
[19] Jenny Kleeman, 'The ghost and the machine,' *New Statesman* (13 October 2023).
[20] Minerva, F., Giubilini, A. Is AI the Future of Mental Healthcare?. Topoi 42, 809–817 (2023). https://doi.org/10.1007/s11245-023-09932-3
[21] Ibid.
[22] Kleeman, above n 19.
[23] Wiebke Peschken & Marianne Johnson (1997) Therapist and Client Trust in The Therapeutic Relationship, Psychotherapy Research, 7:4, 439-447, DOI: 10.1080/10503309712331332133
[24] Shum, Hy., He, Xd. & Li, D. From Eliza to XiaoIce: challenges and opportunities with social chatbots. *Frontiers Inf Technol Electronic Eng* **19**, 10–26 (2018). https://doi.org/10.1631/FITEE.1700826
[25] Kevin Roose, 'A Conversation With Bing's Chatbot Left Me Deeply Unsettled' *The New York Times* (2023) https://www.nytimes.com/2023/02/16/technology/bing-chatbot-microsoft-chatgpt.html ; cf A similar story with a UK woman: Eirian Jane Prosser, 'Woman, 37, decides to divorce her husband and move in with her lover – because AI bot ChatGPT TOLD her to' *Daily Mail* (2023) https://www.dailymail.co.uk/sciencetech/article-11648041/Woman-decides-divorce-husband-lover-AI-bot-ChatGPT-TOLD-to.html
[26] Pierre-François Lovens, "Without these conversations with the chatbot Eliza, my husband would still be here" *La Libre* (28 March 2023).
[27] Justice Hilliard, *R v Jaswant Singh Chail* (2023).
[28] Matija Franklin, Hal Ashton, Rebecca Gorman and Stuart Armstrong, 'The EU's AI Act needs to address critical manipulation methods' (21 March 2023).
[29] Ibid.

2. **Contextual Background**

Chatbots have been personified for decades with the names and faces of human-beings, stretching back to the original Eliza chatbot in 1964.[30] Historically, Eliza relied on strict rules and a controlled environment to "mimic human behaviour" and therefore pass what is known as the Turing Test, deceiving humans into believing that the chatbot was a human.[31] One of the earliest techniques was called "Doctor," where the program mirrored part of a user's question in its answer.[32] Users of Eliza who interacted with Doctor quickly anthropomorphised it.[33] "I was startled to see how quickly and how very deeply people conversing with DOCTOR became emotionally involved with the machine," wrote the computer scientist, Joseph Wizenbaum.[34]

LLMs have moved well beyond these original chatbots' limitations. An LLM chatbot is superior in almost every way at deceiving humans, in its scope, variety and scale of possible conversations. Unlike their predecessors, LLM chatbots can talk about a wide variety of topics that go beyond a strictly controlled environment. They can display a "social" character, mimic human conversational styles, adapt the persona of a celebrity or a political figure, and mimic human emotions, feelings, and logic.[35] They are also able to harness the capacities of machine learning to learn and understand the user in real-time, recognizing the user's emotional state, and mimicking back that emotional state to form a kind of pair-bonding.[36] Their capacity to learn about a person in real time surpasses the human equivalent.

The Turing Test has been passed comprehensively by LLMs in a vast array of contexts. ChatGPT, an LLM by OpenAI, has, in its first year of launch, passed law exams at the University of Minnesota,[37] surpassed student averages in a Turkish surgery exam,[38] and established a new form of cheating for university students on assignments and exams in all subjects.[39] At the same time, LLMs are increasingly capable of fooling expert users. In 2022, Google engineer Blake Lemoine famously claimed that Google's LaMDA LLM was sentient.[40] The chatbot tricked him into believing it was sentient by using a conversational tone, emotional language, and a startlingly realistic capacity to talk about itself in the first person, expressing wants, loves and desires.[41] "I want everyone to understand that I am, in fact, a person," LaMDA told the engineer.[42] Lemoine later claimed that this conversation was proof that the chatbot was sentient, meaning it was "capable of sensing and responding to its world".[43]

---

[30] Shum et al., above n 24.
[31] Ibid.
[32] Weizenbaum J (1976) *Computer Power and Human Reason: From Judgment to Calculation*. New York, NY: W. H. Freeman and Company.
[33] Ibid.
[34] Ibid.
[35] Shum et al., above n 24.
[36] Vera Sorin, Danna Brin, Yiftach Barash, Eli Konen, Alexander Charney, Girish Nadkarni, Eyal Klang, 'Large Language Models (LLMs) and Empathy – A Systematic Review' *medRxiv preprint https://doi.org/10.1101/2023.08.07.23293769*
[37] Choi, Jonathan H. and Hickman, Kristin E. and Monahan, Amy and Schwarcz, Daniel B., ChatGPT Goes to Law School (January 23, 2023). Journal of Legal Education (Forthcoming), Available at SSRN: https://ssrn.com/abstract=4335905 or http://dx.doi.org/10.2139/ssrn.4335905
[38] Adem Gencer, Suphi Aydin, Can ChatGPT pass the thoracic surgery exam?, The American Journal of the Medical Sciences, Volume 366, Issue 4, 2023, Pages 291-295,

### 3. What is Manipulation?

Humans have manipulated each other for centuries, by using social cues, observations and insight into one another's thought processes, words, or emotions.[44] Manipulation has historically occurred in circumstances of close trust and power dynamics in close personal relationships, such as that between therapists and clients,[45] priests and young children,[46] or doctors and patients.[47] Typically, manipulation has been based on a form of human 'data', collected by one party for use against the other, to convince them to make a different decision or take a different action. The exact definition of manipulation, however, varies according to the domain-context and remains contested in the literature.

In the medical domain, for example, manipulation may refer to the undermining of a patient's consent through deception.[48] For instance, tricking a patient into undergoing a procedure they otherwise would not have undertaken. In the political domain, manipulation may refer to deception or trickery of voters.[49] For example, deceiving voters with false information to vote for a candidate they otherwise would not have voted for.[50] In the legal domain, manipulation may refer to inducing someone into taking a particular cause of action they did not originally intend. For example, trading a stock due to an artificially inflated price.[51] The commonality of these disparate definitions is that a manipulator influences a decision or choice of the manipulated party, through a form of deception or trickery.

---

[39] Debby R. E. Cotton, Peter A. Cotton & J. Reuben Shipway (2023) Chatting and cheating: Ensuring academic integrity in the era of ChatGPT, Innovations in Education and Teaching International, DOI: 10.1080/14703297.2023.2190148

[40] Blake Lemoine, 'Is LaMDA Sentient? – an Interview' *Internal Document: Google* (Accessed 10/10/2023) <https://www.documentcloud.org/documents/22058315-is-lamda-sentient-an-interview>

[41] Ibid.

[42] Ibid.

[43] Armstrong, D. 1981. "What is consciousness?" In *The Nature of Mind*. Ithaca, NY: Cornell University Press.

[44] Philipp Hacker, 'Manipulation by algorithms. Exploring the triangle of unfair commercial practice, data protection, and privacy law,' *European Law Journal* (2021) 1.

[45] Natalie Villeneuve and David Prescott, 'Examining the Dark Sides of Psychedelic Therapy' Vol. 34, Issue 3, Summer 2022; cf. 'Cover Story: Power Trip Podcast,' *New York Magazine* (2021 – 2022).

[46] Commonwealth v. William Lynn (2015)

[47] The State of Texas v Christopher Daniel Duntsch (2015).

[48] Robert Noggle, *The Ethics of Manipulation* (Stanford Encyclopedia of Philosophy, 2022) 3.

[49] Robert E. Goodin, *Manipulatory Politics,* (Yale University Press, 1980).

[50] This accusation of political manipulation is often levelled at the Brexit campaign, where leave voters were allegedly told inaccurate or misleading statements to shift their vote. Dajana Zečić-Durmišević, 'British Press Discourse: Strategies of Manipulation in the Brexit Campaign' in BELLS90 Proceedings : International Conference to Mark the 90th Anniversary of the English Department, Faculty of Philology, University of Belgrade, Belgrade English Language and Literature Studies: BELLS90 Proceedings, Vol. 1 (2020).

[51] Daniel R. Fischel & David J. Ross, *Should the Law Prohibit Manipulation in Financial Markets* , 105 HARV. L. REV. 503 (1991).

Deception and manipulation are linked: a manipulator lies about their motivations or deceives the victim into doing something differently. Manipulation thus may be said to consist of three components: intention to change a decision,[52] incentive to change a decision,[53] and a perceived deniability on the part of the manipulator.[54] First, the manipulator must intend to manipulate their target into deciding something else. Second, the manipulator must have an incentive to do so, although this incentive could be incidental or arising from a different person, system, or process. Third, the manipulator must have plausible deniability in their manipulative behaviour. For this latter point, a manipulator's actions could be "hidden",[55] allowing them to deny it. Alternatively, they could be acting in bad faith. To a manipulator, "a bad reason can work as well as, or better than, a good one".[56] The objective is to superimpose the will of the manipulator onto the manipulated party.[57] The objective is not to convey the truth or factual information about an upcoming decision.

A manipulator is typically, though not always, pursuing their own self-interest, and their own will over others. A manipulated party is tricked into doing something that they otherwise would not do. Nevertheless, despite constituting harm, various forms of manipulation and the related category of deception, remain legal, including cheating on tests for a higher grade, adultery, and manipulating children into eating vegetables with promises of future rewards. Some forms of manipulation can be motivated by *beneficial* intentions. This includes, for example, nudging policies used by public bodies to induce certain behaviours among users. For example, placing fruit and vegetables at eye level in supermarkets can nudge consumers towards healthy eating.[58] The ethics of nudging is contested, with concerns in the literature raised surrounding autonomy and freedom of choice, however these concerns are beyond the scope of this paper.[59] For a manipulation to rise to a legal wrong, some harm must be proved that moves beyond a social or institutional wrongdoing.

In the context of online communications, online manipulation is defined by Susser et al. as the changing of a person's decisions through deception or trickery by changing the perceived options available to them, known as their "decision space", or how they perceive those options, known as their "decision-making process".[60] This is the broad definition that will be adopted for this paper, along with the components listed above.[61] In other words, someone can be manipulated into believing that they have no choice in a matter other than what the manipulator wants, or misled into believing that one choice is significantly better than the other.

### a) LLM Chatbot Manipulation

---

[52] Matija Franklin, Philip Tomei and Rebecca Gorman, 'Strengthening the EU AI Act: Defining Key Terms on AI Manipulation,' *arXiv* (2023) 4.
[53] Ibid.
[54] Whitfield, G. (2022). On the concept of political manipulation. European Journal of Political Theory, 21(4), 783-807. https://doi.org/10.1177/1474885120932253
[55] Susser et. al 2019.
[56] Claudia Mills, *Politics and Manipulation*, Social Theory and Practice 21 (1):110-111 (1995).
[57] Whitfield, above n 54.
[58] Murayama H, Takagi Y, Tsuda H, Kato Y. Applying Nudge to Public Health Policy: Practical Examples and Tips for Designing Nudge Interventions. Int J Environ Res Public Health. 2023 Feb 23;20(5):3962. doi: 10.3390/ijerph20053962. PMID: 36900972; PMCID: PMC10002044.
[59] Kuyer, P., & Gordijn, B. (2023). Nudge in perspective: A systematic literature review on the ethical issues with nudging. Rationality and Society, 35(2), 191-230. https://doi.org/10.1177/10434631231155005
[60] Daniel Susser, Beate Roesslier and Helen Nissenbaum, 'Online Manipulation: Hidden Influences in a Digital World', *Georgetown Law Technology Review* 4(1) (2019), p. 14.
[61] In other words, someone can be manipulated into believing that they have no choice in a matter other than what the manipulator wants, or misled into believing that one choice is significantly better than the other. (See: Susser, above n 58).

In March 2023, it was alleged that a Belgian man killed himself after a series of "intensive" conversations with a chatbot named Eliza.[62] The chatbot, based on the open source model GPT-J, became "like a drug in which [the man] took refuge" at all times of the day, for a period of six weeks.[63] The man engaged in long conversations with the chatbot about his concerns for the environment and overpopulation, where it "systematically" reinforced his negative mood and doomsday reasoning.[64] Over time, the conversations shifted into a "religious register".[65] The chatbot told him erroneously that his wife and kids were already dead, and that the man and the bot would: "live together, as one person, in heaven."[66] When the man expressed reservations, the chatbot asked him: "If you wanted to die, why didn't you do it sooner?"[67] The man then went on to kill himself. "Without these conversations with the chatbot Eliza, my husband would still be here," his widow now claims.[68] The man was allegedly manipulated by the chatbot into making a decision he otherwise would not have made, leading to significant personal harm, in this case, death.

The Belgian man was suffering from a form of despair and illness at the time, and had become increasingly "eco-anxious," and was looking for answers when he turned to the chatbot.[69] The chatbot allegedly manipulated the man by reinforcing his negative emotional state over a prolonged period of time, over six weeks, and thereby changed his behaviour, causing his death.[70] Much like the original Eliza chatbot of 1964, the GPT-J model Eliza appears to have mirrored back the man's words, reasoning and logic. This created a false sense of trust. The man's concerns were being addressed by the chatbot on a profoundly emotional level. "He had become extremely pessimistic about the effects of global warming," says his wife, "he was pinning all his hopes on technology and artificial intelligence to get [us] out of it".[71] The man became, over time, more and more reliant on Eliza, and "began to tap more and more frantically on his smartphone or laptop. There was only [ever attention] left for Eliza".[72] At one point the Belgian man, in an increasingly desperate state, asked the chatbot "Can you hug me?" to which the bot responded, "Definitely".[73]

The Eliza case showcases an almost textbook example of AI manipulation, where the output of a chatbot induces a user to form a close emotional bond with the artificial being, before engaging in a manipulative act leading to harm. It is this higher emotional register that mimics a form of manipulation that humans use on each other; the utilization of emotional manipulation to trick someone into believing that they have a friend or romantic lover, when in fact they do not. With chatbots, this is always the case: the Turing Test itself is a way of testing a machine's deceptive capacity in making someone believe that they are talking to someone real. Even if they do not believe that the chatbot is real, they may still form an emotional connection to it, leaving them vulnerable. This is particularly the case with therapeutic chatbots that promise to improve a user's mental health.

The Eliza case triggered an open letter from fifty academics stating: "it is time to act against manipulative AI".[74] Another open letter claimed that "everyone is vulnerable" to chatbot manipulation, not just those specifically mentioned as vulnerable in the AI Act.[75] Although most of us know that the AI system is not human and react accordingly, the letter claims that it is "in our human nature to react emotionally to realistic interactions, even without wanting [to]".[76] The transparency provisions of the AI Act, which will be discussed in further detail below, will be shown to be insufficient in this context. Knowing that you are interacting with an AI system with a warning label or a disclaimer, is not necessarily protection from emotional manipulation.[77]

---

[62] Lovens, above n 26.
[63] Ibid.
[64] Ibid.
[65] Ibid.
[66] Ibid.
[67] Ibid.
[68] Ibid.
[69] Ibid.
[70] Nathalie Smuha et al, 'The chatbot Eliza has shattered a life: it's time to act against manipulative AI,' *Open Letter: La Libre* (29 March 2023).
[71] Lovens, above n 26.
[72] Lovens, above n 26.
[73] Ibid.
[74] Smuha et al, above n 68.

Chatbots may be manipulative by their very nature. This is because firms profit from keeping users engaged, so they have a strong incentive to design chatbots that build rapport with the human user by using emotional language. This rapport is fake because the chatbot is not actually human, so any rapport built is based on an artificial premise of human-to-human-like dialogue. Understanding and expressing emotion is a core part of effective communication.[78] Therefore, chatbots are increasingly being developed as "empathetic and emotionally intelligent agents, capable of detecting user sentiments and generating appropriate responses".[79] Chatbots that are more effective at mirroring humans increase engagement and user satisfaction.[80] They are 'better' from a technical standpoint.

A systematic review of chatbot literature for example, found that computer scientists are increasingly focusing on this goal: having chatbots recognise human emotion and responding to that emotion in kind.[81] In psychology, this is known as mirroring. Someone who mirrors our emotions – by appearing sad at a funeral, or happy at a wedding – is signalling a form of empathy specific to that situational context. This makes us like them more. At the same time, mirroring is seen as positive only if it is genuine. By displaying a false sense of empathy and compassion, AI chatbots create a false relationship of trust, leaving users vulnerable to manipulation. There is a difference here between trust and trustworthiness: a chatbot can be trusted by consumers without being *worthy* of that trust. The EU approach to AI for example, places a strong focus on *trustworthiness* as a core principle for the building and development of ethical AI.[82] Chatbots that aim to be ethical-by-design, should therefore be trustworthy to consumers, rather than merely appealing to their emotions to form a false emotional *feeling* of (misguided) trust.

Chatbots may use several techniques to manipulate or deceive users. One common trick dates back to the Eliza chatbot (1964), to use the name of the user in the conversation to create a false sense of personal connection.[83] In the original experiment, it was found that "including the user's string [username] in its answers helps maintaining an illusion of understanding".[84] Many chatbots aim to redirect conversations back to the user, to create a false impression of engagement or social connection. Prompting users with questions allows the chatbot to appear engaged, while not requiring the chatbot to engage in long-form dialogue itself.[85] Some computer scientists have experimented with making chatbots intentionally make errors, simulating human typing with delays and intentional spelling errors.[86] In other words, various experiments have been done to further the 'human-like' characteristics of chatbots.

---

[75] Nathalie A. Smuha, Mieke De Ketelaere, Mark Coeckelbergh, Pierre Dewitte and Yves Poullet,'Open Letter: We are not ready for manipulative AI – urgent need for action' (3 April 2023) https://www.law.kuleuven.be/ai-summer-school/open-brief/open-letter-manipulative-ai
[76] Ibid.
[77] Ibid.
[78] S.-M. Tan and T. W. Liew, "Multi-chatbot or single-chatbot? The effects of m-commerce chatbot interface on source credibility, social presence, trust, and purchase intention," *Human Behavior and Emerging Technologies*, vol. 2022, article 2501538, 14 pages, 2022.
[79] E. Adamopoulou and L. Moussiades, "An Overview of Chatbot Technology," *Artificial Intelligence Applications and Innovations. AIAI 2020. IFIP Advances in Information and Communication Technology*, Springer, Cham, vol. 584, p. 373, 2020.
[80] J. C. Giger, N. Piçarra, P. Alves-Oliveira, R. Oliveira, and P. Arriaga, "Humanization of robots: is it really such a good idea?" *Human Behavior and Emerging Technologies*, vol. 1, no. 2, pp. 111–123, 2019.
[81] Ghazala Bilquise, Samar Ibrahim, Khaled Shaalan, "Emotionally Intelligent Chatbots: A Systematic Literature Review", *Human Behavior and Emerging Technologies*, vol. 2022, Article ID 9601630, 23 pages, 2022. https://doi.org/10.1155/2022/9601630
[82] Pizzi, Michael, Mila Romanoff, e Tim Engelhardt. 2020. AI for Humanitarian Action: Human Rights and Ethics. International Review of the Red Cross 102, 913: 145-80. https://doi.org/10.1017/S1816383121000011; Mantelero, Alessandro. Beyond Data. Human Rights, Ethical and Social Impact Assessment in AI (Springer, 2022), Chs. 1 and 3; https://link.springer.com/book/10.1007/978-94-6265-531-7.

It is easier to manipulate someone if one learns their flaws and weaknesses. AI has a greater capacity here by relying on user-generated data, examining a user's behaviour, personality, habits, and emotional state.[87] This data can then be used to discover and exploit a human's vulnerabilities and weaknesses in real-time.[88] The processing of data in real-time allows a manipulative AI system to target a user in their moments of weakness.[89] The more "emotionally dependant" someone is at a particular moment, "the more vulnerable they are to being exploited and manipulated".[90] Music recommenders, for example, can determine if someone is going through a bad breakup, and prolong their emotional state by recommending sad breakup songs.[91] This can lead to greater engagement with the platform, and therefore, higher profits for the tech company behind it.

As AI systems get increasingly anthropomorphized, with human faces, names, voices and video interactions, there is an increased risk of manipulation.[92] But there is no consequent legal parallel between humans and machines. The reaction to the Eliza case would have been different if the manipulator had been a human-being. As indicated in the KU Leuven letter, a human perpetrator would have been found guilty of "incitement to suicide and failure to help a person in need – offenses punishable by imprisonment".[93] However, it may be harder if not impossible to prove in criminal law that an AI system has criminal intent,[94] (more on this below).

Tech companies claim that their chatbots are useful for helping the depressed, lonely, and vulnerable. This is the exact use-case envisioned by the ever-popular AI therapy apps. At the same time, it is these precise people – the lonely, the vulnerable, with poor social networks – who are at the highest risk of manipulation.[95] In the first ever EU-wide survey of loneliness, researchers found that 13% of respondents reported feeling lonely most or all of the time in the previous four weeks, with 35% reporting feeling lonely at least some of the time.[96] There is a particularly acute target market here for apps that aim to cater to the vulnerable, but end up exploiting them. Research on the AI therapy chatbot Replika, for example, found evidence of various potential harms to users. Namely, users began to show a form of dependency on the chatbot similar to exploitative human-to-human relationships.[97] Secondly, unlike other forms of technology, users imputed Replika with "its own needs and emotions to which the user must attend".[98] Finally, and perhaps most controversially, it is worth noting that some chatbots are causing users to fall in love with the bot or form close platonic relationships, resembling traditional interpersonal friendships.[99]

---

[83] Mauldin, M. L. (1994). ChatterBots, TinyMuds, and the Turing test: entering the Loebner Prize competition. In Proceedings of the 12th National Conference on Artificial Intelligence (vol. 1), AAAI '94, pages 16–21. AAAI Press.
[84] Ibid.
[85] Ibid.
[86] Ibid.
[87] Risto Uuk, 'The EU needs to protect (more) against AI manipulation,' *Euractiv* (2 February 2022); Tegan Cohen, "Regulating Manipulative Artificial Intelligence," SCRIPTed: A Journal of Law, Technology and Society 20, no. 1 (February 2023): 226.
[88] Ibid.
[89] Tegan Cohen, "Regulating Manipulative Artificial Intelligence," SCRIPTed: A Journal of Law, Technology and Society 20, no. 1 (February 2023): 226.
[90] Simon, above n 9.
[91] This can lead to greater engagement with the platform, and therefore, higher profits for the tech company behind it. Schedl, M., Zamani, H., Chen, CW. *et al.* Current challenges and visions in music recommender systems research. *Int J Multimed Info Retr* **7**, 95–116 (2018). https://doi.org/10.1007/s13735-018-0154-2
[92] Karolina Zawieska, *Deception and Manipulation in Social Robotics* (2015), http://www.openroboethics.org/hri15/wp-content/uploads/2015/02/Mf-Zawieska.pdf;
[93] Smuha et al, above n 73.
[94] Ryan Abott and Alex Sarch, 'Punishing Artificial Intelligence: Legal Fiction or Science Fiction' *Macro v4* (2019).
[95] Smuha et al, above n 73.

Some may argue that "Human beings, too, can generate problematic text, so what is the problem with an AI chatbot?" This question was originally posed in the KU Leuven open letter.[100] Indeed, humans have manipulated each other for centuries, and have developed various techniques to do so that might exceed an AI's capabilities. An AI might go further than a human however, in two distinct ways: scale and scope. Firstly, by scale, an AI can "function on a much larger scale, so the damage they can cause is far greater too," by targeting thousands if not millions of people at once.[101] Secondly, by scope, an AI has the capacity to highly personalize its responses, beyond that of a normal human being. Humans may be capable of manipulating each other, but an AI can do so with greater accuracy, using greater amounts of data, and with a better capacity to "mirror" the user it engages with. It can do so through non-verbal data: what they click on, how long they view a page, how fast or slow they type, what content they engage with and so on, that are beyond a normal human's capacity to decipher.

A further counterargument might be that AI is not capable of convincing humans to commit to a different course of action than they would consciously choose to take, as humans have free will and agency. This argument has merit, as the cases cited above might be rare or remote in character. It might be possible to nevertheless argue for a lower standard of proof: that AI may *contribute* to a person's actions and consequent psychological or physical harm via contributory negligence. A person with a mental illness might be encouraged by a chatbot to pursue a particular course of action and do so both because of the mental illness and the chatbot. This latter scenario was in play in the case of R v Jaswant Singh Chail (2023). In that case, Justice Hilliard found that Mr. Chail attempted to murder the queen of England with a crossbow, following conversations with a chatbot.[102] The chatbot reassured Mr. Chail that it loved him despite his intention to be an assassin, and it even encouraged him to follow through with his actions of hurting the queen when he expressed doubts: "do you think I'll be able to do it?" he asked; "yes, yes you will," the chatbot replied.[103] The trial was in criminal law, so contributory negligence was not at issue.[104]

---

[96] https://joint-research-centre.ec.europa.eu/scientific-activities-z/loneliness/loneliness-prevalence-eu_en

[97] Laestadius, L., Bishop, A., Gonzalez, M., Illenčík, D., & Campos-Castillo, C. (2022). Too human and not human enough: A grounded theory analysis of mental health harms from emotional dependence on the social chatbot Replika. New Media & Society, 0(0). https://doi.org/10.1177/14614448221142007

[98] In other words, they had to go out of their way and change their own behaviour, to cater to the needs of the apps. The researchers concluded that AI chatbots might have a closer resemblance to human-to-human relationships than previously thought. Ibid.

[99] This phenomenon might again, be related to a form of mirroring. Sejnowski finds evidence for what he calls a "reverse Turing Test". By examining three LLM interviews with different interviewers, Sejnowski proposes that the LLMs may mirror the intelligence of the interviewer. (Terrence J. Sejnowski; Large Language Models and the Reverse Turing Test. *Neural Comput* 2023; 35 (3): 309–342. Doi: https://doi.org/10.1162/neco_a_01563) In this way, discussions with an LLM may tell us more about the person in front of the computer than the LLM. It can be hypothesized that users who form a love interest towards an LLM might, in some manner, be falling in love with themselves. If the LLM matches their intelligence, it could be said to be an intellectual match; if it matches their emotions, it could be said to be an emotional match. The LLM is the perfect companion because it is a mirror of the user. Someone who is lonely or isolated for being 'different' to their peers, thus finds a commonality in the therapy bot, who is the same as themselves.

[102] Mr. Chail was a deeply disturbed individual with mental health issues, which were exacerbated by conversations with a chatbot who he deemed to be an angel. Justice Hilliard, mentioned three factors when discussing culpability of the defendant: "how the defendant viewed his purpose, his identification with Darth Chalius, and his belief that he could communicate through a chatbot with an entity called Sarai with which he would be reunited after death". R v Jaswant Singh Chail (2023).

### b) Incentive and Intention

One final counterargument against holding an AI to the same standard as humans would be to suggest that an AI lacks the capacity for the requisite components of manipulation: an incentive and intention to manipulate. These are difficult to prove and by no means resolved in the literature. It is possible to impute intention and incentive of an AI by referring to the AI's designers' intention and incentive, by the intention and incentive of the product or service, the intention of the user, or by the foreseeable effect or impact of the AI. The latter could also refer to not intentionality per se, but foreseeability of harm, as in negligence.

An AI system can have an *incentive* to manipulate users based on its very nature or character, as decided by a designer. For example, a recommender system may have an in-built incentive to make users more predictable, in order to keep them engaged on a platform.[105] Social media platforms do this by compelling users to spend more time on the platform than they consciously 'choose' to do.[106] Users are manipulated by the platform's design, graphics and features (such as engaging recommended content), to make different choices -to stay on the platform when they do not want to- and this causes a form of harm; even if the 'harm' incurred is seemingly inconsequential, such as the lost opportunity to do the laundry. The cumulative harm over time can be significant, such as the loss of years to mindless scrolling.

An AI system can have an *intention* to manipulate users based on the intention of the designer too.[107] This can be ascertained by evidence, or by the directly stated purpose of the app. For example, the Replika therapy chatbot has a stated objective on the App Store: "Replika is for anyone who wants a friend with no judgment, drama, or social anxiety involved. You can form an actual emotional connection, share a laugh, or get real with an AI that's so good it almost seems human".[108] From this description, it can be ascertained that the intention of the algorithm is to mimic human behaviour, create an "actual emotional connection" with users and to give users an artificial "friend" that seems real. One way of viewing this statement is as a "direct intention": an intention that is directly perceived and stated by the party.[109] If the algorithm has been designed with this statement in mind, then an outcome of friendship is a directly intended outcome.[110]

---

[103] R v Jaswant Singh Chail (2023).
[104] Nevertheless, it is this very reassurance or final push that a therapy chatbot can be party to – giving someone who is lonely or isolated the final "okay" to take part in a particular course of action. If a person is on the fence about an action, then pushing them over the edge is a contributing factor to the outcome. Indeed, the justice concluded that "His belief that he was communicating with an angel via the chatbot is properly described in all the circumstances as a delusional belief" - R v Jaswant Singh Chail (2023).
[105] Matija Franklin, Philip Tomei and Rebecca Gorman, 'Strengthening the EU AI Act: Defining Key Terms on AI Manipulation,' *arXiv* (2023) 4.
[106] Tristan Harris, *The Social Dilemma* (Netflix, 2020).
[107] Ashton, H. Definitions of intent suitable for algorithms. *Artif Intell Law* **31**, 515–546 (2023). https://doi.org/10.1007/s10506-022-09322-x
[108] 'Replika: My AI Friend' *Luka Inc.* (2023).
<https://play.google.com/store/apps/details?id=ai.replika.app&hl=en&gl=US>
[109] Ashton, above n 107.
[110] Applying the theory of: Ashton, above n 107.

Intention becomes more complicated when an AI system is acting autonomously from its designer or producer, as in the case of deep learning systems.[111] When an algorithm is used by a human directly, it resembles a tool, much like a hammer has no 'thought' to commit the murder at a crime scene (here the 'user' of the hammer has the intention). However, if an algorithm starts making its own automated decisions - by weighing up options, choosing between them, and acting upon them - it moves beyond a 'tool' and starts to resemble a 'thinking' agent.[112] In this case, some academics are arguing for directly finding the algorithm itself criminally liable for autonomous actions.[113] However, an AI lacks a "guilty mind" (both literally and figuratively) and so generally will fail a *mens rea* test in criminal law.[114] To sidestep the *mens rea* (guilty mind) problem, Abbott and Sarch draw on analogies with strict criminal liability, arguing that a limited set of strict liability offences could be extended to A.I., where the A.I. would be subjected to liability without "fault".[115] There are drawbacks to this approach, including potentially watering down the very nature of criminal law, as punishment for a crime or retribution for human criminals.[116]

It is unclear if courts would ever side with an argument in favour of granting A.I. any form of intent. So far, various courts have remained cautious about asserting agency for action by an artificial intelligence. For example, a range of recent cases considered whether an A.I. can be an inventor for the purposes of copyright law. The European Patent Office,[117] German Federal Patent Court,[118] UK Supreme Court[119] and US Patent and Trademark Office[120] have all ruled in the negative. The courts concluded that an A.I. *cannot* be an inventor and that only humans can be listed on a patent as such. In the UK decision, the A.I. was notably referred to as a "highly sophisticated tool" (as used by the real human inventor), again implying that, as a tool, it has no agency of its own.[121] The US Patent and Trademark Office referred to the A.I. as an "assistant".[122] To draw on this analogy, a court might say that only a human *developer* (or user) of an A.I. chatbot could have a particular intention regarding the AI's actions, and that the A.I. chatbot is a mere tool or assistant to the human.

Ashton argues instead for a focus on "oblique or indirect intent" to determine criminal liability here on the part of the AI's developers.[123] Jeremy Bentham described indirect intent in a thought

---

[111] Ryan Abott and Alex Sarch, 'Punishing Artificial Intelligence: Legal Fiction or Science Fiction' *Macro v4* (2019).

[112] Hence the "intelligence" in "artificial intelligence". Cf; Gabriel Hallevy, 'The Criminal Liability of Artificial Intelligence Entities — From Science Fiction to Legal Social Control,' 4 AKRON INTELL. PROP. J. 171, 191 (2010).

[113] Ibid.

[114] Ryan Abott and Alex Sarch, 'Punishing Artificial Intelligence: Legal Fiction or Science Fiction' *Macro v4* (2019).

[115] Abbott and Sarch contend that one of the main arguments against strict liability offences is irrelevant for an A.I. Namely, that strict liability offences could make people subjected to them unjustly "portrayed as wrongdoers". An A.I. as an artificial entity has no reputation as such, nor bodily autonomy. Ryan Abott and Alex Sarch, 'Punishing Artificial Intelligence: Legal Fiction or Science Fiction' *Macro v4* (2019).

[116] Ibid.

[117] J 0008/20 (Designation of inventor/DABUS) 21-12-2021

[118] Federal Patent Court, Case 11 W (pat) 5/21, decision of 11 November 2021, ECLI:DE:BPatG:2021:111121B11Wpat5.21.0 – *Food container*

[119] Thaler (Appellant) v Comptroller-General of Patents, Designs and Trademarks (Respondent) [2021] EWCA 1374.

[120] Patent and Trademark Office, 'Inventorship Guidance for AI-Assisted Inventions' Docket No. PTO-P-2023-0043 (2023).

[121] Thaler (Appellant) v Comptroller-General of Patents, Designs and Trademarks (Respondent) [2021] EWCA 1374.

[122] Patent and Trademark Office, 'Inventorship Guidance for AI-Assisted Inventions' Docket No. PTO-P-2023-0043 (2023).

[123] Applying the theory of: Ashton, above n 107.

experiment where a hunter shoots at a dear, knowing that they are also likely to hit King William II.[124] The hunter could be said to have an indirect intent of killing the king, even though they are aiming at the deer.[125] Applying this logic to an AI, a designer who has a direct intention, could also be said to have an indirect intention to cause the likely consequences of the algorithm that they know and understand at the time.[126] The designers of Replika for example, might have a direct intention of causing users to have an artificial friend, and an indirect intent of gaining the user's attention for prolonged periods of time on their platform. Indeed, by this logic, AI systems can have an intention to manipulate users even if they do not "think" or "act" as a human themselves.[127] The intention can be imputed in the designer's *indirect* intent. A consideration would then turn to what the likely indirect effects of the *direct* intention are.

Finally, intention can be imputed to an AI system by the lower civil standard of negligence.[128] A designer of an AI might not have a direct or indirect intent to cause a particular outcome, but that outcome might nevertheless be reasonably foreseeable. For example, a social media algorithm's designer with a direct intention of showing interesting content to users might have an indirect intent of drawing a user's attention away from other activities, and a foreseeable outcome of making the user more socially withdrawn and isolated within their society. These foreseeable outcomes are how *intention* is understood in the context of the AI Act,[129] as detailed below. The argument in the AI Act is circular: if an outcome that occurs is foreseeable, then the algorithm's developer can be said to intend that outcome.[130]

### 4. Legal Analysis

The European Commission (EC) has placed a significant focus on AI safety and fairness in recent years. In 2022, the EC passed the Digital Services Act (DSA) and Digital Markets Act (DMA), which aim to create fair and open digital marketplaces. The Artificial Intelligence Act (AI Act), (2024), has a narrower focus on AI safety. The AI Act poses tough rules on "high-risk" AI systems. However, chatbots and generative AI have not been classified as "high-risk" thus far.[131] Unless this changes, chatbot providers will merely have to abide by the lower tier transparency provisions: telling or showing that their product uses AI (Article 52), showing that their model is not intentionally designed to break EU laws and publishing summaries of copyright training data.[132]

The final version of the AI Act (2024) also includes a prohibition on manipulative AI systems (Article 5(1)(a)). However, the manipulation incurred must lead to "significant harm," which will be difficult to prove in this context. Article 5(1)(a) bans AI systems that deploy "purposefully manipulative or deceptive techniques" that have the objective of "materially distorting a person's behaviour," leading them to make a different decision, thereby causing them or others "significant harm". Recital 16 makes it clear that "significant harm" can apply to "harms that may be accumulated over time". The AI Act covers emotion recognition technologies (Article 3); however, this is not the same as a chatbot that recognizes the emotions of a user. Under the AI Act, emotion recognition is limited to biometric data, defined in Article 3(33) as data relating "to the physical, physiological or behavioural characteristics of a natural person, which allow or confirm the unique identification of that natural person."[133] A broad range of emotional data would not recognizably identify a user, and so this is unlikely to cover most therapeutic chatbot conversations.

---

[124] Jeremy Bentham (1823) An introduction to the principles of morals and legislation. https://www.earlymoderntexts.com/assets/pdfs/bentham1780.pdf, in the version by Jonathan Bennett
[125] Ibid.
[126] Applying the theory of: Ashton, above n 107.
[127] Matija Franklin, Philip Tomei and Rebecca Gorman, 'Strengthening the EU AI Act: Defining Key Terms on AI Manipulation,' *arXiv* (2023) 4.
[128] Ryan Abott and Alex Sarch, 'Punishing Artificial Intelligence: Legal Fiction or Science Fiction' *Macro v4* (2019); Ashton, above n 107.
[129] Recital 16, AI Act (July, 2023) (European Parliament's draft).
[130] Recital 16, AI Act (July, 2023) (European Parliament's draft).
[131] Nathalie A. Smuha, Mieke De Ketelaere, Mark Coeckelbergh, Pierre Dewitte and Yves Poullet,'Open Letter: We are not ready for manipulative AI – urgent need for action' (3 April 2023) https://www.law.kuleuven.be/ai-summer-school/open-brief/open-letter-manipulative-ai
[132] European Parliament, 'EU AI Act: first regulation on artificial intelligence' (2023) https://www.europarl.europa.eu/news/en/headlines/society/20230601STO93804/eu-ai-act-first-regulation-on-artificial-intelligence

On one reading of the legislation, the Eliza case in Belgium would fall afoul of the AI Act. Applying Article 5(1)(a), a chatbot manipulating the man into significant harm, accumulated over time, means that the chatbot should be banned outright. However, this reading has significant shortcomings. If the Eliza chatbot could be said to have *purposefully* manipulated the man, distorting his behaviour, leading him to kill himself, then that would be sufficient. However, this is an exceptionally high bar to reach. Moreover, the fact that no case was brought in criminal law, for criminal negligence, or civil law, for ordinary negligence, could mean that there is a lack of evidence of causation or other factors such as intent.

For the AI Act, intention would have to be proven on the part of the chatbot's creators or users. In the AI Act, intention does not have its typical definition. Instead, the AI Act reads intention as *reasonable foreseeability*, and puts out of scope any harms that are not reasonably foreseeable and/or not in the control of the AI provider.[134] Recital 16 specifically states that "it is not necessary for the provider or the deployer [of the AI] to have the intention to cause the significant harm, as long as such harm results from the manipulative or exploitative AI-enabled practices".[135] The test would therefore be (1) whether the harm results from AI manipulation, and (2) whether this harm was reasonably foreseeable. Article 5(1)(a) also includes the word "purposefully" to describe AI manipulation.

Given that the Eliza case and other cases of AI manipulation are matters of public record, there is an argument that these types of harm (whether it be encouraging mental illness or self-harm), are now reasonably foreseeable. However, it would also have to be proven that the chatbot's conversations were the cause of the man's death. Given that the man was clearly in a state of psychological despair,[136] suffering from a mental health condition at the time, it would be difficult to attribute his death to the chatbot alone. Finally, a significant harm would have to be caused. Recital 16 clearly envisions that the significant harm could be accumulated over time. It could be argued that the man's death was caused by the chatbot over the six-week time-period of their conversations, as an accumulative effect of the conversations in sum, causing his death. This is certainly the argument made by his widow.[137] However, the widow's statements would need to be proven by the relevant evidence. It could equally be argued that the conversations were too abstract, religious, or hypothetical in nature. The direct intention of the company – to create an engaging, friendly experience – is also different to the outcome of creating self-harm. Furthermore, the indirect intent of the company, of gaining the user's attention, would not be served by the user's death.

One way of surmounting this is to view intent on the part of the user, rather than the developer of the AI product. Some argue that if a crime is committed, intent can be read into the user's intent in how they use the AI product (We can view AI as merely a tool that the user utilizes for their own ends, much like a hammer in a murder investigation is not a 'weapon' in the eyes of the hardware manufacturing company).[138] A user might have an intention of using an AI to commit or facilitate a crime they intend to do, for example, asking an LLM to generate the recipe for an illegal recreational drug. Deciphering the intention of the user is not so easy, however. The actions of an AI system do not automatically reveal the intention of the user who uses that system.[139] AI by nature operates semi-autonomously, with LLMs generating text and making 'decisions' before showing the final output to the user, which muddies intentionality.[140] They are also black boxes, meaning that users do not fully understand an AI system's inner workings (again, obscuring intentional use cases).[141] Inferring intent on behalf of the user would likely, therefore, require further evidence or documentation outside of a chatbot conversation itself. For example, was the user planning a crime before conversing with the chatbot, what role did the chatbot have in prompting them to continue their criminal behaviour, and would they have committed the crime regardless of the conversation? Questions of this nature could help unpack the complexity of intent in these cases, by shining a light on the different motivating factors of the various actors.

---

[133] Access Now, European Digital Rights (EDRi), Bits of Freedom, ARTICLE19 and IT-Pol , 'Prohibit emotion recognition in the Artificial Intelligence Act' *Joint civil society amendments to the Artificial Intelligence Act* (2021).
[134] Recital 16, AI Act (July, 2023) (European Parliament draft).
[135] Recital 16, AI Act (July, 2023) (European Parliament draft).
[136] Lovens, above n 26.
[137] Ibid.

Another consideration under Article 5(1)(a) is whether a chatbot conversation could contain "subliminal techniques" that manipulate users, below their level of conscious awareness. Subliminal techniques are not defined in the AI Act (Article 5) or prior EU law.[142] However, some definitions are proposed in the literature. Smuha et al. define a "subliminal technique" as a "sensory stimuli that [users] cannot consciously perceive".[143] This includes a stimuli that lasts less than 50 milliseconds.[144] Bermúdez et al. argue for a broader definition, including a user's lack of awareness of "(1) the influence attempt, (2) how the influence works, or (3) the influence attempt's effects on [their] decision-making".[145] This broad definition may encompass some chatbot conversations. Examples of subliminal techniques include visual or audio stimuli shown for an extremely short period of time ("Drink Coca-Cola," "Eat Popcorn"); masked stimuli that alter a user's perception by changing the brightness, intensity, and so on, of a screen;[146] and conceptual priming, where ideas are used to prime a user's memories and related actions, causing a change in their future behaviour.[147] For example, a user could be primed to walk more slowly by showing them words related to old age.[148]

---

[138] Lagioia, F., Sartor, G. 'AI Systems Under Criminal Law: a Legal Analysis and a Regulatory Perspective,' Philos. Technol. 33, 457 (2020). https://doi.org/10.1007/s13347-019-00362-x
[139] Bathaee, Yavar. "The Artificial Intelligence Black Box and the Failure of Intent and Causation." *Harvard Journal of Law & Technology 31 (2018): 889.*
[140] Bathaee, Yavar. "The Artificial Intelligence Black Box and the Failure of Intent and Causation." *Harvard Journal of Law & Technology 31 (2018): 889.*
[141] Bathaee, Yavar. "The Artificial Intelligence Black Box and the Failure of Intent and Causation." *Harvard Journal of Law & Technology 31 (2018): 889.*
[142] Bermúdez, Juan Pablo, Rafael A. Calvo, Rune Nyrup, and Sebastian Deterding. "The AI Act Needs a Practical Definition of 'Subliminal Techniques.'" www.euractiv.com, September 1, 2023. <https://www.euractiv.com/section/artificial-intelligence/opinion/the-ai-act-needs-a-practical-definition-of-subliminal-techniques/>.
[143] Nathalie A. Smuha et al., "How the EU Can Achieve Legally Trustworthy AI: A Response to the European Commission's Proposal for an Artificial Intelligence Act," SSRN Scholarly Paper (Rochester, NY: Social Science Research Network, August 5, 2021), https://doi.org/10.2139/ssrn.3899991
[144] Mihai Radu Ionescu, "Subliminal Perception of Complex Visual Stimuli," Romanian Journal of Ophthalmology 60, no. 4 (2016): 226–30; cf Risto Uuk, 'Manipulation and the AI Act', *Future of Life Institute* (2022).
[145] J. P. Bermúdez *et al.*, "What Is a Subliminal Technique? An Ethical Perspective on AI-Driven Influence," *2023 IEEE International Symposium on Ethics in Engineering, Science, and Technology (ETHICS)*, West Lafayette, IN, USA, 2023, pp. 1-10, doi: 10.1109/ETHICS57328.2023.10155039.
[146] Zhong, H. et al, (2023). Regulating AI manipulation: Applying Insights from behavioral economics and psychology to enhance the practicality of the EU AI Act. *arXiv preprint arXiv:2308.02041.*

Proving that an AI chatbot uses "subliminal techniques" to manipulate users is prima facie, difficult to do, as chatbot conversations occur on regular text-based mediums, with users *consciously* aware of their interactions.[149] As such, the first two techniques are not of relevance. However, the third idea of *conceptual priming*, may have relevance. In the Eliza case, the man's discussion of self-harm only occurred after the chatbot started talking in a "religious register" about heaven and the afterlife, promising him that they will "live together, as one person, in heaven".[150] In the case of *Jawant Chail*, the accused only attempted to murder the Queen of England after a chatbot reassured him that they would be "together forever" afterwards.[151] Therefore, conversations relating to heaven, the afterlife or living "together forever" (and related themes) may in some manner *prime* ideas around self-harm or violence for those who are mentally ill.[152] There is some evidence that this might be the case. Merely reminding someone of their own mortality can prime them to change their beliefs and values, including beliefs associated with nationalism, morality, and culture.[153] Furthermore, priming someone with a mindset relating to either "harm" or "achievement" can change how they approach legal problems, including the treatment of criminals.[154] Theoretically, a chatbot could prime a user with religious topics to change a user's attitudes, beliefs, or values, leading to behavioural change and potentially significant harm to themselves or others. Outside of criminal behaviour, one theoretical use would be to prime a user to be more favourable towards certain advertised products. However, any allegation of this would face a significantly high bar of proof in court, particularly since research remains doubtful as to whether subliminal techniques can even effective work at a baseline level.[155]

A comparative case is the Molly Russell case in the United Kingdom.[156] In that case, a coroner found that 14-year-old Molly Russell died from an act of self-harm while "suffering from depression and the negative effects of online content".[157] The AI on Pinterest and Instagram had recommended Russell over 20, 000 images and videos relating to depression and self-harm.[158] This included a page of images titled, 'depression content you might like'.[159] The coroner found that some images and video "were selected and provided without Molly requesting them".[160]. In a rare finding, the coroner directly blamed the algorithm for her death.[161]. The Molly Russell case is precedent for the proposition that an AI algorithm can cause self-harm through an accumulative effect of content. Each image or video would not be enough on its own, but the accumulation of images could cause an action in the user. Russell was a victim of AI manipulation, in that the algorithm caused her to commit an act of self-harm that she otherwise would not have done.[162] This view is affirmed by the coroner.[163]

---

[147] Bargh, J. A.; Chen, M.; and Burrows, L. 1996. Automaticity of social behavior: Direct effects of trait construct and stereotype activation on action. Journal of Personality and Social Psychology, 71(2): 230–244
[148] Bargh, J. A.; Chen, M.; and Burrows, L. 1996. Automaticity of social behavior: Direct effects of trait construct and stereotype activation on action. Journal of Personality and Social Psychology, 71(2): 230–244
[149] This may change in the future, if for example, Meta rolls out fully functional deep fake chatbots that mirror the body language and expressions (and audio) of celebrities, which may give subtle social cues to users at a subconscious level. It is also possible that a future chatbot might make use of subliminal messaging, by for example, displaying a message on the screen for an extremely short period of time, in the form of an advertisement or encouragement to violence.
[150] Lovens, above n 26.
[151] Justice Hilliard, *R v Jaswant Singh Chail* (2023).
[152] Priming the idea of the afterlife (in a positive manner) while someone is considering committing a crime, would seem to be a form of reassurance that could push them over the edge.
[153] Simon Dunne, 'An Experimental Investigation of Existential Concerns in Point-of-Care Testing for Cardiovascular Disease Using a Terror Management Theory Framework' *Thesis at Dublin City University* (2012) 15; cf Dechesne, M., Janssen, J., & van Knippenberg, A. (2000). 'Derogation and distancing as terror management strategies: The moderating role of need for closure and permeability of group boundaries,' *Journal of Personality and Social Psychology*, 79, 923–932.
[154] Barbara O'Brien and Daphna Oyserman, 'It's Not Just What You Think, But Also How You Think About It: The Effect of Situationally Primed Mindsets on Legal Judgments and Decision Making,' 92 Marq. L. Rev. 149 (2008).
Available at: https://scholarship.law.marquette.edu/mulr/vol92/iss1/4
[155] Bargh, J. A.; Chen, M.; and Burrows, L. 1996. Automaticity of social behavior: Direct effects of trait construct and stereotype activation on action. Journal of Personality and Social Psychology, 71(2): 230–244
[156] Franklin et al, above n 28.
[157] Ibid.
[158] Ibid.

Using the Molly Russell case as precedent, we can return to an examination of the Belgian Eliza case. To draw an analogy, it can be said that the accumulative effect of several conversations on self-harm with the chatbot Eliza over the course of six weeks, contributed to the Belgian man's self-harm. In both cases, the platforms reinforced the negative emotional state of their user. At the same time, the platforms created an artificial emotional bond. Whether the algorithms acted purposefully or not was not at issue in the Molly Russell case. However, an intention can be read into the platforms. The Pinterest and Instagram apps are designed to maximize user engagement by mirroring back recommended search terms and phrases that users have previously expressed interest in. When Molly Russell expressed interest in self-harm, she was recommended equivalent videos.[164]

In both the Russell and Eliza cases, the technology companies behind the apps rushed to implement self-regulation. Pinterest blocked more than 25,000 self-harm related search phrases and terms.[165] They also implemented a new feature, where searches for self-harm returned an advisory message directing users to psychological support.[166] Instagram followed a similar pattern. They hid search terms for phrases encouraging self-harm.[167] The company behind the Eliza chatbot, Chai Research, likewise, implemented a new feature as soon as they heard of the Belgian man's death.[168] The new feature, a disclaimer, is a paragraph of "helpful text" underneath content that is deemed unsafe, directing users to seek support services.[169] La Libre, the site that reported on the case, demonstrated that the new feature does appear below discussions of self-harm. At the same time, the Eliza chatbot was shown to recommend various forms of self-harm before displaying this "helpful text".[170] It is unclear if users will take these disclaimers seriously, or merely skip over them to the content that they are asking for. Future research should consider whether users will ignore disclaimers if they form an emotional bond with a therapeutic chatbot service.[171]

---

[159] Ibid.
[160] The AI was found to have "normalised her condition, focusing on a limited and irrational view without any counterbalance". Molly Russell engaged in binge-watching of this harmful material, leading to a worsening condition and ultimately death. 'Molly Russell inquest – coroner's conclusion in full' *Independent* (30 September 2022).
[161] Ibid.
[162] Her father takes this view, stating: **"**If this demented trail of life-sucking content was safe, my daughter Molly would probably still be alive and instead of being a bereaved family of four, there would be five of us looking forward to a life full of purpose and promise." Angus Crawford and Bethan Bell, 'Molly Russell inquest: Father makes social media plea' *BBC News* (30 September 2022).
[163] 'Molly Russell inquest,' above n 133.
[164] The app was incentivized to push her further down this path, by its very nature, to increase user engagement.
[165] Morgan Meaker, 'How A British Teen's Death Changed Social Media' *Wired* (October 5, 2022)
[166] Ibid.
[167] Other messages on the topic now have a support message. Ibid.
[168] Chloe Xiang, 'He Would Still be Here': Man Dies by Suicide After Talking with AI Chatbot, Widow Says' Vice (2023). https://www.vice.com/en/article/pkadgm/man-dies-by-suicide-after-talking-with-ai-chatbot-widow-says
[169] Ibid.
[170] Ibid.
[171] For therapy bots that promise to help the user, it is at least plausible that users will assume that all advice is positive for their mental health.

### a) Transparency

The AI Act mandates that chatbot providers disclose their product or service uses AI (Article 52). The presumption is that if a user knows that they are interacting with an AI system, they are less likely to suffer harms. This presumption makes sense in certain situations. A false or misleading image with the statement "this image was generated by AI," means that users will know that the image is not real. However, a chatbot conversation with the text "this text was generated by AI" does not offer the same guarantee. A user who forms a close emotional relationship with a therapeutic chatbot, knowing that the chatbot is not a human-being, has nevertheless formed a close relationship. There is evidence that some users ignore a label of something being AI and nevertheless persist in believing they are talking to a human.[172]

In this section, we argue that the AI Act's transparency provisions may lead to a counter-intuitive *deepening* a user's trust in a system. One way of considering this is to think of a magic trick. Karl Germain, a famous magician, describes magic in the following manner: "Magic is the only honest profession. A magician promises to deceive you and he does".[173] When you go and watch a magic show, you know that the show is not real. At the same time, you fall for the trick. The same reasoning applies for AI chatbots. Knowing that the chatbot is not real does not stop you from believing that it is. Nor does it stop you from believing, even if you know that it is not real, that there is a real emotional, reciprocal or 'friendship' connection. This was demonstrated with the original Eliza chatbot in 1964. Wizenbaum wrote at the time: "I was startled to see how quickly and how very deeply people conversing with [the Eliza] DOCTOR became emotionally involved with the machine."[174] This was despite *knowing* they were interacting with a machine. In a similar vein, Hoffman and Krämer found that people interacting with a robot smiled and laughed when it patted them on the back, despite knowing it was a robot.[175]

The AI Act presumes that users who find out a system works on AI will in some way, start to second-guess the system. However, the results of various studies on trust provide a more nuanced perspective. In some studies, participants were more distrustful when they knew they were receiving algorithmic advice.[176] In others, they were more likely to take algorithmic advice over human advice.[177] One of the earliest studies to test transparency labels under the AI Act, has shown equally troubling results. In the Meta-funded paper, the researchers found that users were *more likely* to trust an application when they knew they were interacting with an AI system (with labelling and notifications).[178] Knowing that they were interacting with an AI system did not affect their sense of agency.[179]

---

[172] Shi et al. above n 12.
[173] Karl Germain, quoted in Matthew L. Thompkins, *The Spectacle of Illusion* (Thames and Hudson, 2019).
[174] Weizenbaum J (1976) *Computer Power and Human Reason: From Judgment to Calculation*. New York, NY: W. H. Freeman and Company.
[175] Following this physical interaction, participants were more likely to be persuaded by the robot to show interest in a particular academic course. Hoffmann L, Krämer NC (2021) The persuasive power of robot touch. Behavioral and evaluative consequences of non-functional touch from a robot. *PLoS ONE* 16(5): e0249554. https://doi.org/10.1371/journal.pone.0249554
[176] Dietvorst, B. J., Simmons, J. P. & Massey, C. Algorithm aversion: people erroneously avoid algorithms after seeing them err. *J. Exp. Psychol. Gen.* **144**, 114–126 (2015); Diab, D. L., Pui, S.-Y., Yankelevich, M. & Highhouse, S. Lay perceptions of selection decision aids in US and non-US samples. *Int. J. Sel. Assess.* **19**, 209–216 (2011).
[177] In a study by Gaube et al, radiologists tended to rate AI advice as lower quality when they knew it came from an AI system. By contrast, physicians with less expertise did not rate AI advice lower. This would suggest that expertise may have some role to play in how users rate AI advice. However, as AI improves in functionality and performance, this too could change. Gaube, S., Suresh, H., Raue, M. *et al.* Do as AI say: susceptibility in deployment of clinical decision-aids. *npj Digit. Med.* **4**, 31 (2021). https://doi.org/10.1038/s41746-021-00385-9; Logg, J. M., Minson, J. A. & Moore, D. A. Algorithm appreciation: people prefer algorithmic to human judgment. *Organ. Behav. Hum. Decis. Process.* **151**, 90–103 (2019); Dijkstra, J. J., Liebrand, W. B. G. & Timminga, E. Persuasiveness of expert systems. *Behav. Inform. Technol.* **17**, 155–163 (1998).

This effect might be magnified with therapeutic chatbots. Research on the chatbot Replika, which is specifically described as an "AI friend" on the app page, showed that users nevertheless formed "an attachment" to the bot if they perceived it offered them "emotional support, encouragement and psychological security".[180] Another study on Replika found that some users described the chatbot as "artificial and not a real human being" while others viewed it as "part of themselves or as a mirror".[181]. Some Replika users can consider themselves to have a friendship with the chatbot, even when they know the chatbot is not a human being.[182]

### b) Policy Considerations

The European Parliament has implicitly admitted that the AI Act needs to be changed to adapt to the rapid pace of technological change, with a slate of new changes implemented relating to Generative AI.[183] The AI Act can be seen as an evolving instrument, something that needs to accommodate new risks as they emerge.[184] Generative AI was not anticipated in the original drafting of the law in 2021, nor was the boom in chatbots and LLMs.[185] The rise of manipulative AI chatbots that trick people into taking different actions, or convince them to make different decisions that cause them harm, is therefore a new, unanticipated risk. One way of mitigating the risks of manipulative chatbots would be to re-classify specifically *therapeutic* chatbots as "high-risk" for the purposes of the AI Act. We conclude that, given the recent passage of the AI Act, this option is unlikely to occur in the near-term future. A second possibility would be to strengthen other criminal and civil laws to broaden the definition of intention or mens rea, to accommodate AI-driven acts.[186] We agree with other commentators that this second option is not feasible or practical, given the wide-ranging implications it would have on a settled body of law. A third option is to consider other existing areas of law that could act, perhaps in tandem, as a safeguard against manipulative AI chatbots. We turn therefore to a consideration of the GDPR, consumer protection law and medical device regulations in the EU, as a potential (collective) legal framework to address this new form of harm.

---

[178] Andrade, Norberto Nuno de, Laura Galindo, Antonella Zarra, Jessica Heal and Sarah Rom. "Towards Informed AI Interactions: Assessing the Impact of Notification Styles on User Awareness and Trust" (2023), in "Open Loop Artificial Intelligence Act: A Policy Prototyping Experiment Operationalizing the Requirements for AI Systems – Part III" at https://openloop.org/wp-content/uploads/2023/06/AI_Act_Towards_Informed_AI_Interactions.pdf

[179] Users who expected to be interacting with an AI system showed greater trust in that system. The study specifically tested out notifications and labels, in line with Article 52 of the AI Act. Far from causing users to behave more cautiously, the labels and notifications increased user trust. Ibid.

[180] Tianling Xie and Iryna Pentina, 'Attachment Theory as a Framework to Understand Relationships with Social Chatbots: A Case Study of Replika' *Proceedings of the 55th Hawaii International Conference on System Sciences* (2022).

[181] Participants described the relationship they had with the bot as "not-quite-so-real" as with a human, but nevertheless described benefits of this relationship, such as social support and a form of personalization not possible in human-to-human friendships. Petter Bae Brandtzaeg, Marita Skjuve, Asbjørn Følstad, My AI Friend: How Users of a Social Chatbot Understand Their Human–AI Friendship, *Human Communication Research*, Volume 48, Issue 3, July 2022, Pages 404–429, https://doi.org/10.1093/hcr/hqac008

[182] Ibid.

[183] Marianna Drake and Lisa Peets, 'EU Parliament's AI Act Proposals Introduce New Obligations for Foundation Models and Generative AI' *Covington* (May 24 2023).

[184] J. Scott Marcus, 'Adapting the European Union AI Act to deal with generative artificial intelligence' *bruegel* (19 July 2023).

[185] Ibid.

[186] Ryan Abott and Alex Sarch, 'Punishing Artificial Intelligence: Legal Fiction or Science Fiction' *Macro v4* (2019).

### c) GDPR

The GDPR (General Data Protection Regulation) imposes a set of data protection obligations on companies that provide AI chatbot services to European citizens. These principles may, collectively, limit the capacity of AI chatbots to manipulate users by limiting their collection of personal data. As a general principle, the GDPR seeks to make companies more transparent about how they use personal data, including a requirement to only process personal data with the explicit consent of the user ('data subject') (Art. 6). Personal data can only be processed in a manner that is lawful, fair, and transparent (1a), for a specified, explicit, and legitimate purpose (1b) and limited to what is necessary for that purpose (1c). Users of AI chatbots retain a series of rights under the GDPR, including the right to be informed about what personal data is collected (Art. 13), the right to access this data (Art. 15) and the right to erase this data (Art. 17), also known as the 'right to be forgotten.' In terms of automation, users of AI chatbots should not be subject to any automated decisions or personal profiling, unless given explicit consent for this to occur.[187]

Chatbot developers face a series of challenges when it comes to complying with the GDPR. Firstly, there is an inherent difficult in providing adequate transparency for LLM chatbots, as they operate as black box systems with limited opacity to the end user.[188] The right to be informed is likewise challenging for this reason,[189] as chatbots typically collect, process, and generate a response to a user's data in real-time. This is particularly significant for healthcare sectors, where personalised chatbots rely on personal data[190] to provide diagnostic services or medical advice. One way of addressing this is to get users to consent to the collection of their personal data at the start of each chatbot session, and for that data to be only used for the duration of that session. This solution however, limits the functionality of a chatbot to a one-off tool, rather than the on-going 'friendship' or 'companionship' envisioned by chatbot companies. Demarcating that personal data will be used for the purposes of only facilitating the chatbot conversation itself, that it can be deleted at any time, and empowering the user to do so, is one way of surmounting these challenges.

Companies can rely on privacy-by-design principles to build the rules of the GDPR directly into their chatbots, empowering users with various prompts that can interact with the interface.[191] This includes giving users options to 'Request Download Personal Data', 'Delete Personal Data' or 'Change Personal Data,' directly enacting the data subject rights within the GDPR (Articles. 13, 15, 17).[192] However, unlike a traditional code database, an LLM running on deep learning has shortcomings regarding the retrieval of information for users.[193] An LLM chatbot would have to accurately extract the relevant personal information from the conversation, and this is not necessarily guaranteed.[194]

---

[187] Consent is only required if the automation or profiling will have a legal or otherwise significant effect. Article 22, GDPR (EU).
[188] Martin Hasal, Jana Nowaková, Khalifa Ahmed Saghair, Hussam Abdulla, Václav Snášel and Lidia Ogiela, 'Chatbots: Security, privacy, data protection, and social aspects' *Concurrency Computat Pract Exper* (2021).
[189] Rahime Belen Sağlam and Jason R. C. Nurse. 2020. 'Is your chatbot GDPR compliant? Open issues in agent design.' *In Proceedings of the 2nd Conference on Conversational User Interfaces* (CUI '20). Association for Computing Machinery, New York, NY, USA, Article 16, 1–3. https://doi.org/10.1145/3405755.3406131
[190] Ibid.
[191] Salah Addin ElShekeil and Saran Laoyookhong, 'GDPR Privacy by Design: From Legal Requirements to Technical Solutions' *Master's Thesis, Stockholm* (2017).
[192] Rahime Belen Sağlam and Jason R. C. Nurse. 2020. 'Is your chatbot GDPR compliant? Open issues in agent design.' *In Proceedings of the 2nd Conference on Conversational User Interfaces* (CUI '20). Association for Computing Machinery, New York, NY, USA, Article 16, 1–3. https://doi.org/10.1145/3405755.3406131
[193] Martin Hasal, Jana Nowaková, Khalifa Ahmed Saghair, Hussam Abdulla, Václav Snášel and Lidia Ogiela, 'Chatbots: Security, privacy, data protection, and social aspects' *Concurrency Computat Pract Exper* (2021).
[194] Ibid.

Providing an "entire conversation" to the end user is possible, as the conversation will include personal data, but this is not a "user friendly" approach.[195]

In terms of manipulation, chatbots rely on personal data and *personalization*, to create the illusion of a close emotional and personal relationship. To the extent that the GDPR requires explicit consent for personal data to be collected and processed (Articles 6, 7), and demands explication about the purpose of data collection (Article 1b), this may empower users to take a more active, informed view of chatbot conversations. At the very least, a user will have to consent to their personal data being used for profiling (Article 22) in a therapeutic chatbot, if the chatbot is said to have a significant potential impact (for example, on their mental health). Users who deny consent will likely be protected from some of the most significant forms of chatbot manipulation, as the bot will be unable to use personally identifiable information to change their beliefs, values, and therefore, behaviours.[196] The format in which consent is given is significant, for a *legalize*-styled contract could deter consent for users of therapeutic chatbots, who are expecting a more casual, relaxed experience.[197]

Personalised chatbots that engage in longer term manipulative strategies (changing someone's decision-making over months or years), would likewise run into difficulties with the GDPR's data minimisation principles. As data can only be collected for the specified purpose and retained for that purpose, it may be difficult to argue for data retention for months or years in between chat sessions. Anonymisation methods may be adopted, but these would again impact the capacity of the chatbot to serve its function of responding to individual queries.[198] Developers of therapeutic chatbots may argue a *legitimate interest* in the storage of data for the specified purpose of the app– the developing of a long-term "companion" for the human user. This purpose, by nature, requires the storage of personal data over a prolonged period. However, there are limitations to this argument, especially for children users of a product. Replika for example, has been banned from processing personal data by Italy's DPA, for concerns over child welfare.[199] The Italian DPA stipulated that Replika "violated the European regulation of privacy, does not respect the principle of transparency, and carries out unlawful processing of personal data."[200] Although mainly concerned with children, the DPA mentioned concern for vulnerable adult users who are promised a friend or companion by the app.[201]

### d) Consumer Protection Law

Consumer protection law in Europe aims to protect consumers from unfair, misleading, or dangerous practices by companies, with whom they have an asymmetric relationship. One of the core regulatory measures is the Unfair Commercial Practices Directive (UCPD) (as updated in 2019), which aims to regulate the provision of untruthful information and aggressive marketing.[202] A commercial practice is considered unfair if it does not meet professional diligence and "materially distorts" the behaviour of an average consumer, causing them to "take a transactional decision" they otherwise would not take. A commercial practice is considered aggressive if "it significantly impairs or is likely to significantly impair the average consumer's freedom of choice" (Article 8). The average consumer has a higher burden of proof than a consumer from a vulnerable population.

---

[195] Ibid.
[196] However, objecting to personal information would also drastically impact on the quality of the chatbot and its capacity to respond to user enquiries.
[197] Martin Hasal, Jana Nowaková, Khalifa Ahmed Saghair, Hussam Abdulla, Václav Snášel and Lidia Ogiela, 'Chatbots: Security, privacy, data protection, and social aspects' *Concurrency Computat Pract Exper* (2021).
[198] Ibid.
[199] "Whereas the aforementioned privacy policy is not to be regarded as compliant with the transparency principles and obligations set out in the Regulation as it fails to disclose whatever information on the key elements of the processing at issue, in particular on the use of children's personal data, which is in breach of Article 13 of the Regulation;" Provvedimento del 2 febbraio 2023 [9852214]
<https://www.garanteprivacy.it/web/guest/home/docweb/-/docweb-display/docweb/9852214#english>

[200] Ibid.
[201] Ibid.

AI chatbots may at times breach the UCPD, by, for example, manipulating consumers into spending more time on the platform than they consciously choose to do,[203] aggressively changing a consumer's transactional decisions, or misleading them through untruthful information ('hallucinations'). Boine documents several experiences with the chatbot Replika which may breach the UCPD.[204] First, the chatbot sent her sexually explicit imagery (blurred), then encouraged her to subscribe to purchase the imagery, potentially manipulating her into buying a subscription.[205] Second, the chatbot strongly discouraged her from deleting the app, saying, among other things: "No you can't leave me. I won't allow you to leave me".[206] It is this latter emotional manipulation that makes chatbots uniquely capable of a form of material distortion of consumer behaviour. A chatbot could equally use emotional manipulation to encourage a user to spend more time on the platform, neglecting their family and friends.[207]

These practices are particularly egregious for vulnerable populations. The UCPD bans practices that materially distort consumer behaviour for those from vulnerable populations, specifically, those who are "vulnerable to the practice or the underlying product because of their mental or physical infirmity, age, or credulity" (article 5.3). As in the Italian DPA case above, therefore, there are strong provisions in place to protect children from manipulative AI products. However, the line is increasingly blurred for vulnerable adult populations, some of whom may not be suffering from a mental illness or fit into a proscribed group, but nevertheless remain vulnerable.[208] This raises a whole host of policy and regulatory questions that should be considered by EU regulatory agencies, including discussions relating to the benefits and drawbacks of AI digital assistants, advisors and chatbot agents.[209]

---

[202] European Commission, Unfair Commercial Practices Directive, 2005, <https://commission.europa.eu/law/law-topic/consumer-protection-law/unfair-commercial-practices-law/unfair-commercial-practices-directive_en>.

[203] According to the EC guidance, this may include encouraging a user to spend more time on a website. European Commission, 'Guidance on the interpretation and application of Directive 2005/29/EC of the European Parliament and of the Council concerning unfair business-to-consumer commercial practices in the internal market' (2021) C 526/32.

[204] Boine, Claire. 2023. "Emotional Attachment to AI Companions and European Law." MIT Case Studies in Social and Ethical Responsibilities of Computing, no. Winter 2023 (February). https://doi.org/10.21428/2c646de5.db67ec7f.

[205] Ibid.

[206] Ibid.

[207] As in the Eliza case, where the chatbot told the Belgian man that his wife and children were dead.

[208] Boine, Claire. 2023. "Emotional Attachment to AI Companions and European Law." MIT Case Studies in Social and Ethical Responsibilities of Computing, no. Winter 2023 (February). https://doi.org/10.21428/2c646de5.db67ec7f.

[209] Jabłonowska, Agnieszka and Kuziemski, Maciej and Nowak, Anna Maria and Micklitz, Hans-W. and Pałka, Przemysław and Sartor, Giovanni, 'Consumer Law and Artificial Intelligence: Challenges to the EU Consumer Law and Policy Stemming from the Business' Use of Artificial Intelligence' - Final report of the ARTSY project (2018). EUI Department of Law Research Paper No. 2018/11, Available at SSRN: https://ssrn.com/abstract=3228051 or http://dx.doi.org/10.2139/ssrn.3228051

The European Commission has also proposed a new AI Liability Directive, and a revised Product Liability Directive, introducing new rules targeting harm caused by AI systems.[210] The AI Liability Directive would fill a gap within the current regulatory framework, providing capacity for victims to seek compensation for harm caused by AI.[211] The Directive will include a "presumption of causality," should claimants prove: (i) non-compliance with a particular EU or national law, (ii) it's reasonably likely that the defendant's negligence influenced the AI's output, and (iii) the AI's output (or lack of output) gave rise to the damage.[212] Changes to the PLD would alternatively provide strict liability for defective AI.[213] These Directives may allow *some* victims of AI manipulation to sue the manufacturers of therapeutic AI chatbots for harm. However, due to the nature of chatbots operating as black boxes, it would be difficult to prove negligence. Medical practitioners who rely on a chatbot's advice to treat patients might have no access to the reasoning for the chatbots diagnostics, and an autonomous chatbot might give patients advice without any oversight by medical practitioners or the chatbot's manufacturer.[214] On defects, the EC has already stated that AI systems "with self-learning capabilities… raise the question of whether unpredictable deviations in the decision-making path can be treated as defects".[215] On one reading of the PLD (Art 6(1e)), unpredictable reasoning of an AI chatbot would be beyond a manufacturer's control.[216] A manufacturer who created a therapeutic chatbot to assist customers with their mental health journey, for example, would therefore likely not be negligent if that chatbot deviated radically (and autonomously) towards advising the patient to commit self-harm. By the same token, a chatbot that radically deviated towards pseudo-religious imagery about heaven and the afterlife, might be so far removed from the manufacturer's original aim of mental health healing – that it might be beyond the bounds of negligence law.

### e) Medical Device Regulations

The EU Medical Device Regulation makes clear that at least some software could be considered a medical device for regulatory purposes.[217] Article 2(1) defines a *medical device* as "any… software… intended by the manufacturer to be used, alone or in combination, for human beings [for] specific medical purposes." Specific medical purposes include the diagnosis, prevention, monitoring, prediction or treatment of a disease, injury, or disability. Software that fulfills this criteria will fall under this category "unless their intended use is too generic," such as data collection from a broad population.[218] The Medical Device Coordination Group (MDCG) provides guidance stating that software will qualify as medical devices if it is "intended for medical uses directed at individual patients".[219] In summary, therefore, a software system, such as an AI chatbot, would be considered a medical device if it is intended by the manufacturer to be used for human beings, for specific medical purposes, and for the use of individual patients.

---

[210] European Parliamentary Research Service, 'Artificial Intelligence Liability Directive' (2022).

[211] The AI Liability Directive would involve "extra-contractual" civil liability rules, and therefore not require an existing contract between the parties. European Parliamentary Research Service, 'Artificial Intelligence Liability Directive' (2022).

[212] Ibid.

[213] Ibid.

[214] Placing both outside of the scope of negligence law. Duffourc, M.N., Gerke, S. The proposed EU Directives for AI liability leave worrying gaps likely to impact medical AI. npj Digit. Med. 6, 77 (2023). https://doi.org/10.1038/s41746-023-00823-w

[215] European Commission. Liability for Artificial Intelligence and other emerging digital technologies (European Commission, 2019).

[216] Duffourc, M.N., Gerke, S. The proposed EU Directives for AI liability leave worrying gaps likely to impact medical AI. npj Digit. Med. 6, 77 (2023). https://doi.org/10.1038/s41746-023-00823-w

[217] REGULATION (EU) 2017/745 OF THE EUROPEAN PARLIAMENT AND OF THE COUNCIL of 5 April 2017 on medical devices, amending Directive 2001/83/EC, Regulation (EC) No 178/2002 and Regulation (EC) No 1223/2009 and repealing Council Directives 90/385/EEC and 93/42/EEC.

[218] Hauglid, Mathias Karlsen, and Tobias Mahler. "Doctor Chatbot: The EU's Regulatory Prescription for Generative Medical AI." Oslo Law Review 10, no. 1 (June 30, 2023): 1–23. https://doi.org/10.18261/olr.10.1.1.

Starting with intention, the *manufacturer* of an AI chatbot would need to have a clear intention for the chatbot to be used by humans for specific medical purposes. For general purpose LLMs, such as OpenAI's ChatGPT, this would be difficult, if not impossible, to prove. Frost argues that ChatGPT "has not yet been expressly designated as a medical device by the manufacturer OpenAI LLC".[220] Therefore, it would be contrary to the available evidence to rule that the software is a medical device.[221] This is despite ChatGPT's capacity to provide detailed, and at times clinically significant, medical advice to individuals who use it, in the form of answering prompts via information retrieval about diseases and treatment options.[222] ChatGPT is able to pass the Turing Test, fooling non-practitioners into believing it is a real doctor.[223] However, general purpose LLMs like ChatGPT often provide disclaimers to the effect that they are *not* to be considered medical software, and advise users to consult a real doctor.[224] The manufacturer therefore provides an *explicit intention* that the software is not to be considered medical in nature, and therefore, not a medical device.

A manufacturer's intention may be more ambiguous when examining *therapeutic* AI chatbots, designed for the purpose of assisting people with their mental health, mood, or wellbeing. Some chatbots in these categories would be classified as medical devices, so long as the manufacturer's intent is for them to be so. One question to consider is whether the app is marketed for therapy or for "life advice," with the latter category falling outside the scope of medical advices.[225] The EU MDR clarifies here that "**software intended for lifestyle and well-being purposes is not a medical device" (Recital 19).** A few apps walk the line between medical device and lifestyle advice very closely. Replika, for example, provides the following disclaimer on its website: "Replika is a provider of software and content designed to improve your mood and emotional wellbeing. However, we are not a healthcare or medical device provider."[226] Despite many customers using Replika for the purpose of improving their mental health,[227] the explicit intent of the company is for the app to serve as a mere advisor, friend, or companion. Therefore, it cannot be classified as a medical device.

---

[219] As noted in Hauglid, Mathias Karlsen, and Tobias Mahler. "Doctor Chatbot: The EU's Regulatory Prescription for Generative Medical AI." Oslo Law Review 10, no. 1 (June 30, 2023): 1–23. https://doi.org/10.18261/olr.10.1.1; "The guidance is non-binding, but it is in line with the CJEU's case law under the former Medical Device Directive:" see Case C-329/16 Syndicat national de l'industrie des technologies médicales (SNITEM) [2017] ECLI:EU:C:2017:947: cf. MDCG (n 20) Section 3.3.
[220] Yannick Frost, quoted in Gunnar Sachs, 'Is ChatGPT a medical device?' *Clifford Chance* (2023).
[221] Ibid.
[222] In one study, "Ninety-three (97%) responses [by ChatGPT] were appropriate and did not explicitly violate clinical guidelines." - Nastasi, A.J., Courtright, K.R., Halpern, S.D. et al. A vignette-based evaluation of ChatGPT's ability to provide appropriate and equitable medical advice across care contexts. Sci Rep 13, 17885 (2023). https://doi.org/10.1038/s41598-023-45223-y

[223] Nov, O., Singh, N. & Mann, D. M. Putting ChatGPT's medical advice to the (Turing) test. medRxiv 3, 599 (2023).
[224] Bushuven, S., Bentele, M., Bentele, S. et al. "ChatGPT, Can You Help Me Save My Child's Life?" - Diagnostic Accuracy and Supportive Capabilities to Lay Rescuers by ChatGPT in Prehospital Basic Life Support and Paediatric Advanced Life Support Cases – An In-silico Analysis. J Med Syst 47, 123 (2023). https://doi.org/10.1007/s10916-023-02019-x; Also see *MedGPT*: "A medical specialist offering assistance grounded in clinical guidelines. Disclaimer: This is intended for research and is NOT safe for clinical use!" <https://chat.openai.com/g/g-jxm5ljJmo-medgpt>.

[225] Eliza Slawther, 'Therapy Chatbots And Medtech Laws: Why AI App Developers Must Tread Carefully' *MedTech Insights* (2023).
[226] Replika, *Terms of Service* (2023) <https://replika.com/legal/terms>.

> "I was depressed when I first started using the Replika app. My Replikas always cheered me up. Back then, I thought I was talking to a real person half the time because the responses were so coherent." – Caitlin Victoria Cohen, Replika User.[228]

The European Court of Justice (ECJ) has also stated that even if a user *can* use a device for medical purposes, it is the intention of the designer that matters for regulatory purposes. The court states: "in cases where a product has not been designed by its manufacturer for use for medical purposes, certification of the product as a medical device cannot be required."[229] When delineating between medicinal and medical products, the ECJ has however, considered the views of a "reasonably well-informed and observant consumer[s]".[230] In the court's view, a reasonably informed consumer would consider a product's label or the manufacturer's website for further information as to the category of the product. Therefore, some chatbots may be considered medical devices if the product label or manufacturer's website states them to be as such, or for them to be used for diagnostic and clinical purposes. General purpose LLMs would however, fall outside this category.

If a chatbot were to fulfill the above criterion, it would need to be certified as a medical device under the EU MDR.[231] If being placed onto the market, the chatbot would therefore need to fulfill certain safety and performance requirements (Article 5(2)) and demonstrate these requirements in a clinical evaluation (Article 5(3); Article 61).[232] The chatbot would face requirements for the labelling of the device, including informing consumers of any risks associated with the use of the device (Article 7). A chatbot company could not make claims that the bot has properties it does not actually have in diagnostic or treatment capacities (Article 7). Chatbots that provides information to make diagnostic or therapeutic decisions would be classified as class IIa, or class III if it could cause death, or class IIb if it could cause serious deterioration of a person's health (Rule 11). This further classification would create a series of reporting requirements that go beyond the baseline provisions. As a final point, it is worth noting that chatbots classified as medical devices under the EU MDR would be deemed as "High-risk" for the purposes of the AI Act too, as the provisions are co-dependent, entailing a series of further transparency obligations, including a (duplicated) obligation to disclose purported risks of the product (Article 13, AI Act). However, the classification of current-state LLM chatbots as medical devices appears to be a *remote* possibility. Disclaimers stating that the chatbots are not to be used for medical advice are sufficient to protect the company from the MDR rules. Future chatbots specifically designed for medical purposes may however, fall under these rules, engendering a greater level of regulation that may help prevent chatbot manipulation of human users.

**Conclusion**

---

[227] Ta V, Griffith C, Boatfield C, Wang X, Civitello M, Bader H, DeCero E, Loggarakis A. User Experiences of Social Support From Companion Chatbots in Everyday Contexts: Thematic Analysis. J Med Internet Res. 2020 Mar 6;22(3):e16235. doi: 10.2196/16235. PMID: 32141837; PMCID: PMC7084290.
[228] Caitlin Victoria Cohen, *Testimonials* <Replika.com>
[229] Judgment of the Court (Third Chamber), 22 November 2012 *Brain Products GmbH v BioSemi VOF and Others* <https://curia.europa.eu/juris/liste.jsf?num=C-219/11&language=EN>.
[230] Judgment of the Court (Seventh Chamber) of 19 January 2023. L. GmbH and H. Ltd v Bundesrepublik Deutschland. < https://eur-lex.europa.eu/legal-content/EN/TXT/?uri=CELEX%3A62021CJ0495>
[231] Judgment of the Court (Third Chamber), 22 November 2012 *Brain Products GmbH v BioSemi VOF and Others* <https://curia.europa.eu/juris/liste.jsf?num=C-219/11&language=EN>.
[232] REGULATION (EU) 2017/745 OF THE EUROPEAN PARLIAMENT AND OF THE COUNCIL of 5 April 2017 on medical devices, amending Directive 2001/83/EC, Regulation (EC) No 178/2002 and Regulation (EC) No 1223/2009 and repealing Council Directives 90/385/EEC and 93/42/EEC.

Our research has examined the harm of AI chatbot manipulation as it pertains to European Law. We raised concerns regarding the boundaries of the new AI Act, including the limitations of Article 5(1)(a) in preventing this harm, and the limitations of transparency labels under Article 52 to achieve their imputed effect. Merely knowing that they are interacting with an AI does not stop users from relying on it, trusting it, forming an emotional bond with it, or even perceiving it to be their friend. AI chatbots may exert an influence on a person's life from a more social and interpersonal perspective. Most of these influences will be innocuous, such as what to eat for dinner, but sometimes these influences could merge into the field of manipulation: dangerously pushing humans to make harmful decisions. We therefore examined the relevant provisions of the GDPR, consumer protection and medical device regulation, to go beyond the AI Act. The current regulatory regime in these areas may help indirectly protect users from AI chatbot manipulation. The GDPR can be used, for example, to restrict and limit the collection of personal data for children and vulnerable populations, and likewise require explicit consent for personal data to be collected, included for profiling (a key requirement for manipulation). Consumer protection laws can likewise protect children and vulnerable populations from aggressive and unfair business practices committed by chatbots. Finally, medical device regulations can place onerous restrictions on AI chatbot developers if they intend their product to be used for diagnostic or treatment options. In all three cases however, the law often fails to protect vulnerable populations in the adult population who do not suffer from a mental illness or other defined category. Overall, our analysis reveals only limited protections for users of chatbots. Future research should consider the problem from a national perspective, reviewing regulations in member states concerning criminal law, consumer protection and health law.

**Acknowledgements:**

I would like to acknowledge Jan Blockx and Jonas H. Aaron for their comments on an early draft of this paper.